\documentclass[aps,prl,reprint,superscriptaddress]{revtex4-1} 

\usepackage{graphicx} 
\usepackage{natbib} 
\usepackage[usenames,dvipsnames]{color} 
\usepackage{amsmath} 
\usepackage{amssymb} 
\usepackage{soul} 
\usepackage{ifthen} 

\newcommand{\nn}{\nonumber}

\newcommand{\ket}[1]{|{#1}\rangle}

\def\l{\left}
\def\r{\right}
\newcommand{\sch}{Schr\"odinger~}
\def\be#1\ee{\begin{equation}#1\end{equation}}
\def\ba#1\ea{\begin{align}#1\end{align}}
\def\bg#1\eg{\begin{gather}#1\end{gather}}
\def\t{\text}

\def\eq#1{(\ref{eq:#1})}

\def\shownote{1} 
\newcommand{\note}[1]{\ifthenelse{\shownote=1}{\textcolor{Red}{[[#1]]}}{}}
\def\showaddmat{1} 
\newcommand{\addmat}[1]{\ifthenelse{\showaddmat=1}{\textcolor{Gray}{[[#1]]}}{}}

\begin{document}

\title{Observation of Floquet States in a Strongly Driven Artificial Atom}

\author{Chunqing Deng}
\email{cdeng@uwaterloo.ca}
\affiliation{Institute for Quantum Computing, Department of Physics and Astronomy, and Waterloo Institute for Nanotechnology, University of Waterloo, Waterloo, Ontario, Canada N2L 3G1}

\author{Jean-Luc Orgiazzi}
\affiliation{Institute for Quantum Computing, Department of Electrical and Computer Engineering, and Waterloo Institute for Nanotechnology, University of Waterloo, Waterloo, Ontario, Canada N2L 3G1}

\author{Feiruo Shen}
\affiliation{Institute for Quantum Computing, Department of Physics and Astronomy, and Waterloo Institute for Nanotechnology, University of Waterloo, Waterloo, Ontario, Canada N2L 3G1}

\author{Sahel Ashhab}
\affiliation{Qatar Environment and Energy Research Institute (QEERI), HBKU, Qatar Foundation, Doha, Qatar}

\author{Adrian Lupascu}
\affiliation{Institute for Quantum Computing, Department of Physics and Astronomy, and Waterloo Institute for Nanotechnology, University of Waterloo, Waterloo, Ontario, Canada N2L 3G1}

\date{ \today}

\begin{abstract}
We present experiments on the driven dynamics of a two-level superconducting artificial atom. The driving strength reaches 4.78~GHz, significantly exceeding the transition frequency of 2.288~GHz. The observed dynamics is described in terms of quasienergies and quasienergy states, in agreement with Floquet theory. In addition, we observe the role of pulse shaping in the dynamics, as determined by nonadiabatic transitions between Floquet states, and we implement subnanosecond single-qubit operations. These results pave the way to quantum control using strong driving with applications in quantum technologies. \end{abstract}

\maketitle

Monochromatic driving is the most common tool in quantum control, applicable to various physical systems including nuclear and electronic spins, atoms, ions, superconducting qubits, and quantum dots~\cite{vandersypen_2004_revNMR}. For driving that is weak compared to the relevant transition frequency, the dynamics can be described in terms of Rabi oscillations between energy eigenstates. In contrast, with strong driving the commonly used rotating wave approximation~\cite{Bloch_1940_BSshift} breaks down, resulting in complex evolution. Strong driving dynamics is most adequately described in the framework of Floquet theory~\cite{Shirley:1965sI}, where the state of a driven system is expressed in terms of quasienergies and quasienergy states. Exploring this more general framework expands the field of quantum control, gaining increasing relevance as current experiments on the implementation of high-fidelity quantum gates~\cite{Gustavsson:2013go} and protection against decoherence~\cite{Yoshihara:2014db} are performed with a driving strength that is a significant fraction of the transition frequency. In addition, strong driving is relevant in the fields of quantum sensing, for phase measurements~\cite{griffith_1998_phaseandrisetimelasers}, and quantum simulation, for designing effective Hamiltonians in the emerging field of Floquet engineering~\cite{goldman_2014_floquetengineering}.

In this Letter we report experiments on the dynamics of an artificial atom, a superconducting quantum bit~\cite{Devoret:2013jz, Nori:2011jj}, strongly driven by a microwave field. Strong driving has been studied in the field of atomic physics, using either optical~\cite{paulus_2003_IonizationvsLaserPhase} or radio-frequency pulses~\cite{griffith_1998_phaseandrisetimelasers}. In experiments with NV centers in diamond, time dynamics was observed for driving strength up to values comparable to the transition frequency~\cite{Fuchs:2009ca}. Superconducting qubits display a naturally strong coupling to electromagnetic fields due to their mesoscopic character. Previous experiments on strong driving of superconducting qubits have mostly addressed the steady-state response to continuous waves~\cite{saito_2004_multiphoton,Oliver:2005rK, Sillanpaeae_2006_LZCPB, Wilson:2007dd, izmalkov_2008_coupledqubits, sun_2009_LZ, Tuorila2010, Silveri:2014t_, shytov_2003_interferometry,Ashhab:2007ep, Shevchenko:2010hf}. A few experiments have observed time-domain Rabi oscillations~\cite{Nakamura:2001cr, Chiorescu2004, Saito2006} with a driving strength exceeding the transition frequency, and two of these demonstrated good agreement with the theoretically predicted Bessel-function dependence of the Rabi frequency~\cite{Nakamura:2001cr,Saito2006}. In our experiments, we use quantum state tomography to investigate the dynamics of a superconducting qubit strongly driven by microwave pulses with controllable shape. The observed system dynamics is very well described in terms of quasienergies and quasienergy states, as predicted by Floquet theory. In particular we observe several frequency components in the dynamics, in very good agreement with theory. We find that the switching on and off of the driving pulse plays an important role in the qubit evolution, as determined by adiabaticity conditions in the Floquet picture~\cite{Drese:1999tq}. We also used strong driving for fast, subnanosecond, preparation of qubit states.

\begin{figure}[!]
\includegraphics[width=86mm]{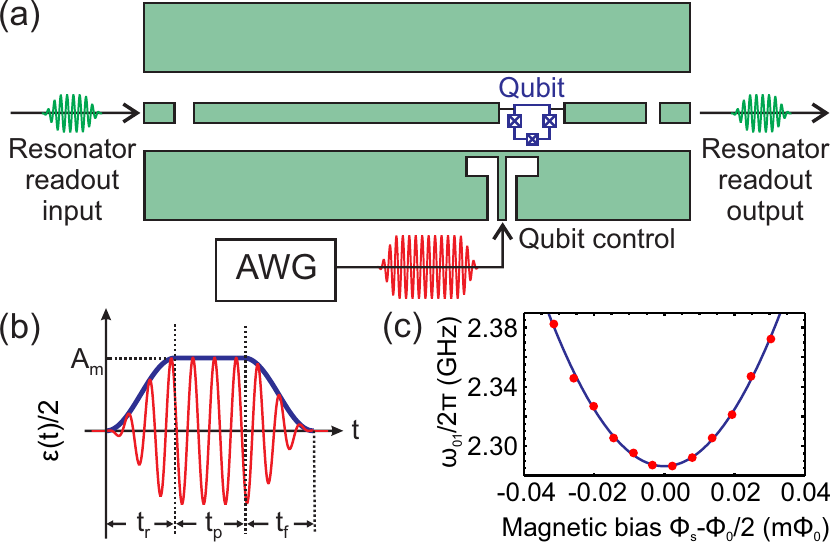}
\caption{\label{fig1}(a) Schematic representation of the experimental setup. The qubit, formed by a superconducting loop interrupted by Josephson junctions (cross symbols), is coupled to a superconducting coplanar waveguide resonator. Readout is based on the transmission of a microwave pulse from the resonator input (left) to its output port (right). A waveguide (bottom) is used to couple microwave control pulses to the qubit. (b) Representation of qubit control pulses, with rise and fall times $t_\t{r}$ and $t_\t{f}$ respectively, and maximum-amplitude duration $t_\t{p}$. The thick line indicates the pulse envelope, which reaches a maximum $A_\t{m}$. During rise and fall, the envelope is shaped as $\frac{A_\t{m}}{2}\l(1-\cos\l(\pi t/t_\t{r}\r)\r)$ and $\frac{A_\t{m}}{2}\l(1+\cos\l(\pi (t-t_\t{p}-t_\t{r})/t_\t{f}\r)\r)$. (c) Qubit transition frequency $\omega_{01}$ from spectroscopy measurements versus the static magnetic flux $\Phi_\t{s}$. The continuous line is a fit of the transition frequency, yielding the parameters $\Delta=2\pi\times 2.288$~GHz and $I_p=690$~nA.}
\end{figure}

The artificial atom in our experiment is a superconducting flux qubit~\cite{Mooij:1999sQ}. Among the different types of superconducting qubits, flux qubits have the advantage of high-level anharmonicity, leading to ideal two-level system behavior, and of strong coupling to electromagnetic fields~\cite{Niemczyk:2010gv, Forn:2010by} which enable strong driving. Qubit state measurement is performed by probing microwave transmission through a resonator coupled to the qubit (see Fig.~\ref{fig1}(a)), in the dispersive regime of circuit quantum electrodynamics~\cite{wallraff_2005_1}.

The qubit Hamiltonian is given by $H(t)=-\frac{\hbar \Delta}{2} \sigma_z -\frac{\hbar \epsilon(t)}{2} \sigma_x$ in a basis formed by symmetric and antisymmetric combinations of clockwise and anticlockwise persistent current states in the qubit loop~\cite{Mooij:1999sQ}. Here $\Delta$, the minimum energy level splitting, is a fixed parameter, and $\epsilon (t) = 2I_p(\Phi(t) -\Phi_0/2)$, with $\Phi(t)$ the magnetic flux applied to the loop dependent on the time $t$. The magnetic flux $\Phi(t) = \Phi_\t{s} + \Phi_d(t)$ with $\Phi_\t{s}$ a static flux generated by a superconducting coil and $\Phi_d(t)$ a time-varying magnetic flux coupled to the qubit through a waveguide terminated by an antenna (see Fig.~\ref{fig1}(a)). The usual approach employed to generate control pulses is based on using a modulator to shape the quadratures of a continuous wave produced by a frequency synthesizer. Here we use instead a new generation of high-speed arbitrary waveform generator (AWG) to directly synthesize the microwave pulses~\cite{supple}, leading to the time accuracy required for control with subnanosecond resolution pulses.

A plot of the qubit transition frequency versus the static flux $\Phi_\t{s}$, obtained by spectroscopy with weak and long microwave pulses, is shown in Fig.~\ref{fig1}(c). All the experiments reported in this Letter are performed at the symmetry point ($\Phi_\t{s} = \Phi_0/2$), where the qubit transition frequency $\omega_{01}=\Delta = 2\pi\times 2.288$~GHz. We use amplitude shaped pulses $\epsilon (t) = 2 A(t) \cos (\omega t)$, with $A(t)$ characterized by a maximum amplitude $A_\t{m}$ and rise and fall times denoted by $t_\t{r}$ and $t_\t{f}$ respectively (see Fig.~\ref{fig1}(b)). At the symmetry point, the energy relaxation and pure dephasing times are given by $T_1=1.8$~$\mu$s and $T_\t{Ramsey}=0.3$~$\mu$s. These coherence times, currently limited by quasiparticle tunneling, microscopic two-level systems, and charge noise~\cite{Orgiazzi:2014wo, Stern:2014hh}, can be further improved by infrared shielding techniques and improved qubit design without impairing the ability to strongly drive the qubit.

\begin{figure*}[!]
\includegraphics[width=180mm]{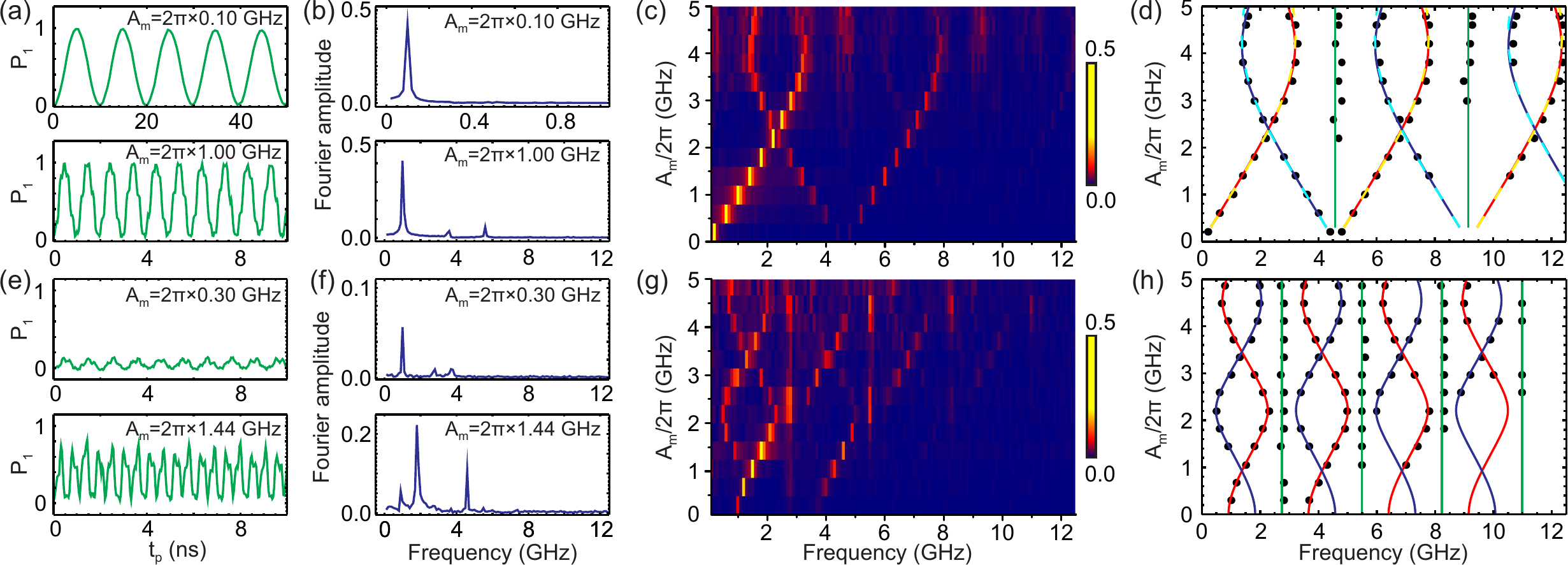}
\caption{\label{fig2}Coherent oscillations versus driving amplitude for resonant driving ($\omega=\Delta$, (a-d)) and off-resonance driving ($\omega=0.6\times\Delta$, (e-h)). (a,e) Qubit excited state probability $P_1$ versus control pulse duration $t_\t{p}$ for $A_\t{m}=0.10$ and 1.00~GHz (a) and $A_\t{m}=0.30$ and 1.44~GHz (e). (b,f) Discrete Fourier transforms of the oscillations in (a) and (e) respectively. (c,g) Color plot of the Fourier transform of population oscillations versus frequency and driving pulse amplitude. (d,h) Positions of peaks in the Fourier transform of coherent oscillations versus driving amplitude extracted from the data in (c) and (g) (dots). The lines are plots of $n\omega$, $n\omega-\Delta \epsilon$, and $n\omega+\Delta \epsilon$ respectively, with the quasienergy difference $\Delta \epsilon$ determined numerically and $n$ an even integer. For the resonant driving case (d) we show in addition corresponding curves (dashed) based on the analytical approximation for $\Delta \epsilon$ discussed in the text (see also Ref.~\cite{supple}).}
\end{figure*}

Experiments are performed by repeating, typically 16,384 times, a sequence formed of state reset, control using an applied pulse, and measurement in the energy eigenbasis. Figure~\ref{fig2}(a) shows the qubit's average excited state probability versus the duration of the microwave pulse, with driving on resonance. The waveform is defined with zero rise and fall times; however, the actual rise and fall times are determined by the analog bandwidth of the AWG and are specified to be shorter than 22 ps~\cite{supple}. For weak driving (Fig.~\ref{fig2}(a), top panel), sinusoidal oscillations are obtained, as predicted based on the rotating wave approximation. With a large Rabi driving strength (Fig.~\ref{fig2}(a), bottom panel), large amplitude oscillations are accompanied by smaller amplitude faster oscillations, a signature of non-negligible counter-rotating term effects. The different frequency components are clearly visible in the Fourier transform of the signal (Fig.~\ref{fig2}(b)). For a wide range of the driving strength $A_\t{m}$, from 2$\pi\times$0.20~GHz to 2$\pi\times$4.78~GHz, the Fourier transformed data are shown in Fig.~\ref{fig2}(c).

The presence of the various frequency components in the Rabi oscillations can be understood based on Floquet theory, which predicts that for a time-periodic Hamiltonian with period $T$ the quantum state is given by $\ket{\psi(t)}= \sum_{j=0,1} c_j e^{-i\epsilon_j t}\ket{u_j(t)}$ with $\epsilon_j$ the quasienergies and $\ket{u_j(t)}$ the quasienergy states, periodic in time with period $T$. As a result, the probability to find the system in its excited state is expected to show oscillatory behavior with frequency components $n \omega$ and $\pm \Delta\epsilon + n \omega$, with $\Delta \epsilon$ the quasienergy difference, $\omega=2\pi/T$ the driving frequency, and $n$ any integer number. The harmonic drive signal used in our experiment has the additional symmetry $\epsilon(t+T/2)=-\epsilon(t)$, and as a result only components with even $n$ values are present~\cite{Creffield_2003_FloquetCrossings}. Fig.~\ref{fig2}(d) shows the extracted frequency components versus driving amplitude. We compare the experimental results with calculations of the quasienergies based on numerical simulations (solid lines) and an analytical expression (dashed lines). The latter, obtained based on approximate diagonalization after transformation to a rotating frame~\cite{supple}, gives a quasienergy difference $\Delta\epsilon = \omega \sqrt{\l( 1 - J_0\l( \frac{2A}{\omega} \r) \r)^2 + J_1^2\l( \frac{2A}{\omega}\r) }$, with $J_{0/1}$ Bessel functions of the first kind and order 0/1. This formula provides a good approximation for the case of a two-level system biased at its symmetry point and driven on or near resonance with arbitrary strength, complementing previous theoretical work where the weak- and strong-driving limits of this formula had been derived~\cite{Ashhab:2007ep,Shevchenko:2010hf}. Additional experiments were performed with the qubit driven off-resonance, with a driving frequency $\omega=2\pi \times 1.373$~GHz (see Fig.~\ref{fig2}(e) and (f)). The Fourier transform of the qubit population signal and the identified frequency components are shown in Figs.~\ref{fig2}(g) and (h) respectively. Good agreement with the predictions of numerical calculations is observed in this case as well (see Fig.~\ref{fig2}(h)).

\begin{figure}[!]
\includegraphics[width=86mm]{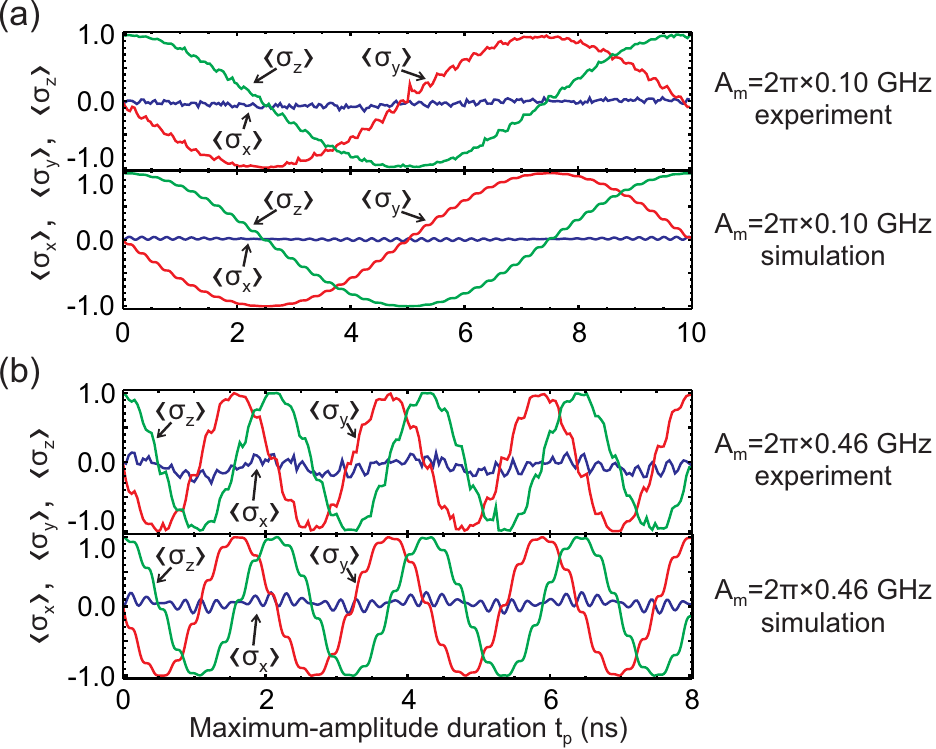}
\caption{\label{fig3}Measurements and simulations of the evolution of the Bloch vector components, given by the average values of the Pauli $\sigma_\alpha$ ($\alpha = x,y,x$) operators, after a pulse with zero rise and fall time, versus the length of the pulse for $A_\t{m} = 2\pi\times$0.10~GHz (a) and $A_\t{m} = 2\pi\times$0.46~GHz (b). The experimental results are in excellent agreement with the results of numerical simulations.}
\end{figure}

Tomography experiments confirm the role of the counter-rotating terms in the driven evolution of the qubit. Fig.~\ref{fig3} shows results of state tomography versus the duration of driving pulses for two values of the driving amplitude, $A_\t{m} = 2\pi\times0.10$~GHz and $A_\t{m}=2\pi\times0.46$~GHz, and zero rise and fall times. For both values of the driving amplitude, high-amplitude oscillations are observed, with a period corresponding to the quasienergy difference. In the weak-driving limit, these oscillations are the usual Rabi oscillations. High-frequency components are observed in addition, with a significant amplitude at strong driving, reflecting the presence of the non-negligible counter-rotating wave component. The results of tomography experiments are in very good agreement with predictions of numerical simulations of the \sch equation (see Fig.~\ref{fig3}).

\begin{figure*}[!]
\includegraphics[width=180mm]{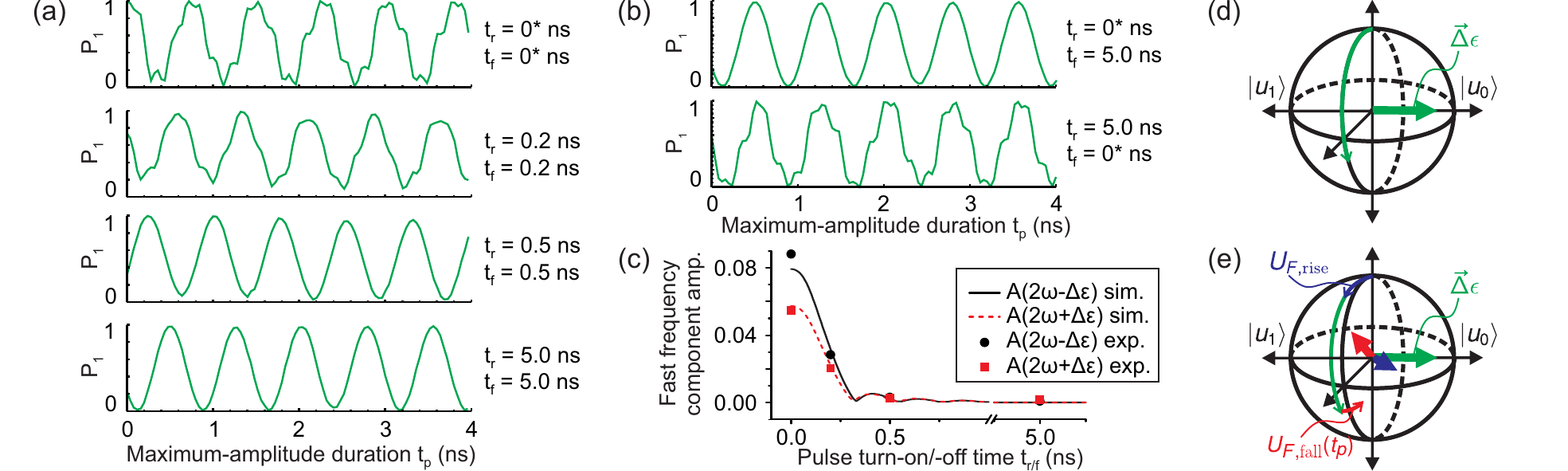}
\caption{\label{fig4}(a,b) Qubit excited state probability $P_1$ versus pulse duration $t_\t{p}$ for various rise (fall) times $t_\t{r}$ ($t_\t{f}$) and equal maximum amplitude $A_\t{m}=2\pi\times 1.33$~GHz. Panels (a) and (b) show data with symmetric or asymmetric rise and fall. (c) The measured (dots and squares) and simulated (continuous and dashed lines) fast oscillation amplitudes, at frequencies $2\omega-\Delta\epsilon$ and $2\omega+\Delta\epsilon$ respectively. (d,e) Adiabatic (d) and nonadiabatic (e) evolution in the Floquet picture. The state is represented on the Bloch sphere, with a basis chosen such that the instantaneous quasienergy states $\ket{u_{0,1}}$ are in the equatorial plane. The state vector evolution (thin arrows) is a rotation around a fictitious field (thick arrows). The initial qubit state is $(\ket{u_0}+\ket{u_1})/\sqrt{2}$. In the adiabatic case (d), the state evolution is described by a phase $-\int_0^t \Delta\epsilon(t) \t{d}t$ applied to $\ket{u_{1}}$, which is equivalent to rotation around the fictitious field $\vec\Delta\epsilon(t)$ aligned with $\ket{u_0}$. In this picture, the evolution of the qubit is a simple rotation, although in the energy eigenbasis the qubit undergoes complex dynamics. In the nonadiabatic case (e), transitions between Floquet states arise during pulse turn-on and turn-off, characterized by the unitary transformations $U_{F,\t{rise}}$ and $U_{F,\t{fall}}$ respectively, which correspond in general to rotations around axes that are not parallel to $\vec\Delta\epsilon(t)$. The asterisk (*) indicates that the actual rise or fall time is around 20~ps, determined by the analog bandwidth of the AWG.}
\end{figure*}

The presence of the fast oscillatory terms in the driven evolution depends not only on the pulse amplitude, but also on the pulse turn-on and turn-off times. Fig.~\ref{fig4}(a) shows qubit state oscillations for a driving strength $A_\t{m}=2\pi\times 1.33$~GHz, and different rise and fall times. Fast oscillatory terms are gradually suppressed as the turn-on and turn-off times are increased. We emphasize that fast oscillatory components in the oscillations are completely suppressed for slow pulse turn-on and turn-off despite the fact that during most of the driven evolution the driving amplitude is comparable with the transition frequency. The absence of fast oscillations for slow turn-on and turn-off can be understood based on adiabaticity in the Floquet picture~\cite{Drese:1999tq}. Indeed, the time-dependent qubit state can be written, up to an overall phase and a geometric phase, as $\ket{\psi(t)} = c_0(t)\ket{u_0(A,t)}+c_1(t)e^{-i\int_0^t \Delta\epsilon(t) \t{d}t}\ket{u_1(A, t)}$, with $\ket{u_0(A,t)}$ and $\ket{u_1(A,t)}$ the instantaneous driving-amplitude-dependent quasienergy states. The initial values of the coefficients $c_0$ and $c_1$ are determined by the representation of the initial qubit state, which is the ground state, in the basis formed by the states $\ket{u_{0,1}(0,0)}=(\ket{0}\pm \ket{1})/\sqrt{2}$, with $\ket{0}$($\ket{1}$) the ground/excited state of the qubit~\cite{supple}. For slowly varying driving amplitude $A(t)$, the evolution is adiabatic in the Floquet basis, and therefore the coefficients $c_0$ and $c_1$ maintain their initial values. The dynamics of the qubit in this case, using the Bloch sphere representation (see Fig.~\ref{fig4}(d)), can be understood by the rotation of the pseudospin representing the state around a fictitious field determined at any given time by the difference $\Delta\epsilon$ between the quasienergies. Similarly to the weak-driving case, the qubit simply undergoes a Rabi rotation between the ground and excited states, with a rotation angle given by $1/2\int_0^t \Delta\epsilon(t) \t{d}t$.

With short rise and fall times, the evolution of the qubit at the beginning and end of the pulse is nonadiabatic in the Floquet representation. Nonadiabatic effects can be described by unitary transformations $U_{F,\t{rise}}$ and $U_{F,\t{fall}}$ at the beginning and the end of the pulse respectively (see Fig.~\ref{fig4}(e)). The latter depends periodically on the pulse duration with period $T$, leading to fast oscillations of the final state of the qubit. For a driving amplitude $A_\t{m}=2\pi\times 1.33$~GHz, the qubit population dynamics is well described by a sum of oscillatory terms at frequencies $\Delta \epsilon$, $2\omega + \Delta \epsilon$, and $2\omega -\Delta \epsilon$. In Fig.~\ref{fig4}(c) we plot the amplitude of the high-frequency components, at $2\omega \pm \Delta \epsilon$, versus the pulse rise and fall time. The experimental results are in good agreement with values extracted based on numerical simulations of the qubit evolution.

In additional experiments (see Fig.~\ref{fig4}(b)), we observed the evolution of the qubit with strong pulses and asymmetric rise and fall times. The final state of the qubit displays fast oscillations for pulses with slow rise and fast fall, whereas fast oscillations are absent for pulses with fast rise and slow fall. This observation confirms the asymmetric role of the two rotations, $U_{F,\t{rise}}$ and $U_{F,\t{fall}}$.

We next discuss the use of strong driving for fast quantum gates, specifically qubit state preparation. Starting with the qubit in its ground state, we apply pulses with a driving strength $A_\t{m}=2\pi\times 0.46\,\t{GHz}$ and rise and fall times of approximately $20$~ps, defined by the AWG bandwidth. The state $\l(\ket{0}-i\ket{1}\r)/\sqrt{2}$ is prepared in $0.48$~ns with a fidelity of $0.9996 \pm 0.0006$~\cite{supple}. Similarly, state $\ket{1}$ is prepared in $1.08$~ns with a fidelity of $0.9969 \pm 0.0008$. We have performed numerical simulations of state evolution, which predict state preparation fidelities of 0.9997 and 0.9976 for states $\l(\ket{0}-i\ket{1}\r)/\sqrt{2}$ and $\ket{1}$ respectively, in good agreement with the experimental results. Future work should address the optimization of gate fidelities based on randomized benchmarking~\cite{Gustavsson:2013go, Barends2014, Sheldon2015}.

Our work demonstrates the feasibility of using strong driving for the control of superconducting artificial atoms. The dynamics was analyzed in the framework of Floquet theory. The consideration of adiabaticity in the Floquet picture provides a valuable viewpoint on dynamics, applicable well beyond the regime where the rotating wave approximation holds. Our experimental demonstration brings very exciting prospects for experiments addressing the interplay between Floquet dynamics and environmental effects~\cite{Grifoni_1998_dqt,Moskalets_2002_FloquetQuantumPumps}. We expect that our results will stimulate new work across a broad range of fields including  quantum computing, open system dynamics, quantum simulation, and quantum sensing.

\begin{acknowledgments}
We thank Martin Otto, Ali Yurtalan, and Feyruz Kitapli for help with the experiments and the members of the University of Waterloo Quantum Nanofab team for assistance on device fabrication. We would like to thank Kevin Resch and Sergey Shevchenko for comments on the manuscript. We are very grateful to Mark Skadorwa from Tektronix for valuable discussions of technical specifications and facilitating the lending of the AWG used in the experiments. We acknowledge support from NSERC, Canada Foundation for Innovation, Ontario Ministry of Research and Innovation, Industry Canada, and the Canadian Microelectronics Corporation. During this work, CD was supported by an Ontario Graduate Scholarship and AL was supported by an Early Research Award.
\end{acknowledgments}

%

\pagebreak
\widetext
\begin{center}
\textbf{\large Supplemental Material}
\end{center}
\setcounter{equation}{0}
\setcounter{figure}{0}
\setcounter{table}{0}
\setcounter{page}{1}
\makeatletter
\renewcommand{\theequation}{S\arabic{equation}}
\renewcommand{\thefigure}{S\arabic{figure}}
\renewcommand{\bibnumfmt}[1]{[S#1]}
\renewcommand{\citenumfont}[1]{S#1}


\section*{experimental METHODS}
\subsection*{Sample fabrication and parameters}
\begin{figure}[b]
\includegraphics[width=120mm]{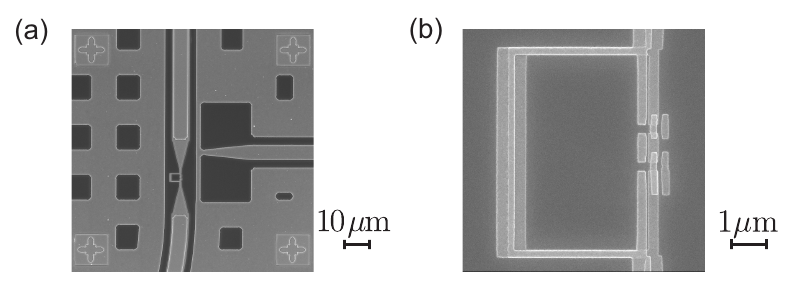}
\caption{(a) Scanning electron microscope (SEM) image showing a qubit embedded in the resonator. An antenna (right) is used to couple microwave control pulses to the qubit. (b) SEM image of a qubit device nominally identical to that used in this work.} \label{fig0}
\end{figure}
The device (see Fig.~\ref{fig0} for images) is fabricated on a high-resistivity silicon substrate, in two steps. Firstly, the resonator and the control lines are defined by optical lithography, followed by evaporation of a 190 nm thick aluminum layer and liftoff. In the second step, a bilayer resist is patterned by electron-beam lithography. After an argon milling step, shadow evaporation of two aluminum layers, 40 and 65 nm thick respectively, followed by liftoff, defines the qubit. The resonator is formed by a coplanar waveguide interrupted by coupling capacitors at the input and output ports. The fundamental half-wavelength mode used for qubit readout has a resonance frequency $\omega_\t{r}=2\pi\times 6.641\,\t{GHz}$. The qubit and resonator are strongly coupled using the inductance of a shared line, with a coupling strength for the fundamental mode $g=2\pi\times 537\,\t{MHz}$~\cite{S_abdumalikov_2009_flux-qubit,S_Bal_2012_QubitDetector}.

\subsection*{Measurement setup}
The experiments are performed in a dilution refrigerator at a temperature of 35~mK. The fabricated device is enclosed in a copper box, which is placed inside a three-layer high-permeability metal shield. An active magnetic field compensation system is used to further reduce the effect of fluctuations of the external magnetic field. The device is connected to room temperature electronics using coaxial cables, which include attenuators and filters placed at different temperature stages. The signal at the output port of the resonator is amplified using a low-noise high electron mobility transistor amplifier with a noise temperature of 4~K. The qubit control pulses are directly synthesized by an arbitrary waveform generator Tektronix AWG70002A at a sampling rate of 25~GS/s. The AWG has an analog bandwidth of 13.5~GHz and an intrinsic rise/fall time of less than 22~ps.

\subsection*{Pulse calibration for quantum state tomography}
The quantum state tomography experiments are performed by applying pre-rotation pulses to the qubit before the readout. We use pre-rotation pulses denoted by unitary operations including identity $I$ and $\pi/2$ rotations around the $x$- and $y$-axis, denoted by $R_x(\pi/2)$ and $R_y(\pi/2)$ respectively. The pre-rotation pulses $R_x(\pi/2)$ and $R_y(\pi/2)$ are defined with $t_\t{r}=t_\t{f} = 0.2$~ns and $A_\t{m} = 2\pi\times 130$~MHz, where the dynamics are well described by the RWA. The length and the phase of these pulses are chosen so that their rotation angles, as well as the angle between their rotation axes, are calibrated to $\pi/2$. To calibrate the rotation angles and the rotation axes of these pulses, we manipulate the qubit with a pulse sequence consisting of the pre-rotation pulses and then measure the final state of the qubit. For calibrating the rotation angle $\theta$ of pulse $R_{x(y)}(\theta)$, we use a pulse sequence corresponding to a unitary operation $\l[R_{x(y)}\l(\theta\r)\r]^{2n+1}$ which amplifies the rotation angle error by $2n+1$ times and projects this angle error to the measurement basis. Using the above pulse sequence with $n=5$, we determine an optimized pulse length of $t_\t{p}=1.9$~ns and an upper bound of 0.003~rad on the error of the rotation angles around both axes. For calibrating the rotation axes, we use a pulse sequence corresponding to unitary operation $R_{y'}\l(\frac{\pi}{2}\r) \l\{ \l[R_{x}\l(\frac{\pi}{2}\r)\r]^{2} \l[ R_{y'}\l(\frac{\pi}{2}\r)\r]^{2} \r\}^{n} R_{x}\l(\frac{\pi}{2}\r)$, where $R_{y'}\l(\frac{\pi}{2}\r)$ denotes the rotation around the axis $y'$ to be calibrated. Using pulse sequence with $n=5$ which amplifies the error in the rotation axis by $2n$ times, we determine an upper bound of 0.002~rad on the error of the rotation axis $y'$. These rotation-angle and rotation-axis errors lead to an error less than $0.003$ for each Bloch vector component $\sigma_i$. These errors are much smaller than the statistical errors from the measurements which will be discussed in the next section.

\subsection*{Quantum state preparation with strong pulses}
The density matrices of states are reconstructed from the quantum state tomography data using maximum likelihood estimation. A direct comparison between the reconstructed state $\rho$ from experimental results and the ideal state $\rho_\t{ideal}$ is given by the state fidelity $F = \t{Tr}\l(\sqrt{\sqrt\rho_\t{ideal} \rho \sqrt\rho_\t{ideal}} \r)$. Statistical errors of the reconstructed states and their fidelities are determined by the parametric bootstrapping method~\cite{S_EfronTibshirani199405}. The procedure is described as follows: 1. Calculations of the standard deviation $\sigma$ of the Bloch vector $\vec\sigma = (\sigma_x, \sigma_y, \sigma_z)$ obtained from the tomography experiments using the measurement statistics. According to the central limit theorem, the average of the random variables $\bar\sigma_i = \sum_n \sigma_{i,n}/n$ ($i=x,y,z$) has a normal distribution with a mean of $\langle\sigma_i\rangle$ and a standard deviation of $\sigma(\sigma_i)/\sqrt{n}$ with $n$ the number of repeated experiments. 2. Generation of random data of the Bloch vector from the normal distributions above. The size $B$ of the random data need to be large, usually $B>100$. 3. For each random data, the maximum likelihood estimation is used to estimate the quantum state $\rho$ and the state fidelity $F$ with respect to the target state is calculated. Finally, statistics for all the data of size $B$ is collected. The variance of data serves as an estimator of the standard error.

Using pulses with $A_\t{m} = 2\pi\times 0.46$~GHz and rise and fall times about $20$~ps, we prepare states targeting $\ket{-Y}=(\ket{0}-i\ket{1})/\sqrt{2}$ and $\ket{1}$ from the ground state $\ket{0}$ with a total pulse length of $0.48$~ns and $1.08$~ns respectively. The reconstructed density matrices of the actual states are determined to be:
\bg
\rho_{\ket{-Y}} = \begin{pmatrix}
                   0.511048  & -0.0145217 + 0.499667i \\
                   -0.0145217 - 0.499667i & 0.488952  \\
                 \end{pmatrix}, \nn \\
\rho_{\ket{1}} = \begin{pmatrix}
                   0.00590452 & -0.0709229 + 0.0289758i \\
                   -0.0709229 - 0.0289758i & 0.994095 \\
                 \end{pmatrix}, \nn
\eg
of which the fidelities are $0.9996 \pm 0.0006$ and $0.9969 \pm 0.0008$ respectively.

\section*{Calculation of quasienergies and quasienergy states}
\subsection*{Matrix form of the Floquet Hamiltonian}
Here, we derive a matrix form for the Floquet Hamiltonian of a qubit under harmonic driving for numerical calculation of the quasienergies. In the derivation below, we shall generally follow similar steps to those given in Ref.~\cite{S_Son:2009eg}.

We consider a two-level artificial atom under harmonic driving. The Hamiltonian is given by:
\begin{equation}
H = - \frac{\Delta}{2} \sigma_z + A \cos \left( \omega t \right) \sigma_x.
\end{equation}
We consider units with $\hbar = 1$ from here onwards. In order to simplify the appearance of expressions that will appear in the derivation below, we make a basis transformation described by the operation
\begin{equation}
\psi_{\rm rot} = e^{-i\frac{\pi}{4}\sigma_y} \psi,
\end{equation}
i.e.~a $\pi/2$ rotation about the y axis. The Hamiltonian is then transformed into the form
\begin{equation}
H_{\rm rot} = - \frac{\Delta}{2} \sigma_x - A \cos \left( \omega t \right) \sigma_z.
\end{equation}
Since we are dealing with a two-dimensional Hilbert space, the time-periodic Hamiltonian gives rise to two Floquet states, which can be expressed as
\begin{equation}
\ket{\psi_{F,j}(t)} = e^{-i\epsilon_jt} \ket{u_j(t)},
\end{equation}
where $\epsilon_j$ are the quasienergies and $\ket{u_j(t)}$ are the periodic components in the Floquet states, \emph{i.e.} time-dependent quantum states with period $T=2\pi/\omega$:
\begin{equation}
\ket{u_j(t)} = \sum_{n=-\infty}^{\infty} e^{in\omega t} \ket{u_{j,n}}.
\end{equation}
Substituting the above expression in the Schr\"odinger equation, we find the relation
\begin{equation}
\epsilon_j \ket{u_{j,n}} = \left( - \frac{\Delta}{2} \sigma_x + n\omega \right) \ket{u_{j,n}} - \frac{A}{2} \sigma_z \left( \ket{u_{j,n-1}} + \ket{u_{j,n+1}} \right)
\end{equation}
The above set of equations can be expressed as a single equation:
\begin{equation}
\epsilon_j \ket{U_j} = H_F \ket{U_j} \label{eq:eigenequation}
\end{equation}
where $U_j$ is the vector $\{\dots , u_{j,n-1,\uparrow} , u_{j,n-1,\downarrow} , u_{j,n,\uparrow} , u_{j,n,\downarrow} , u_{j,n+1,\uparrow} , u_{j,n+1,\downarrow} , \dots \}$, and the Floquet Hamiltonian $H_F$ is given by
\begin{equation}
H_F = \left(
\begin{array}{cccccccc}
\ddots & & & & & & & \\
& (n-1) \omega & \displaystyle{-\frac{\Delta}{2}} & \displaystyle{-\frac{A}{2}} & 0 & 0 & 0 & \\
& \displaystyle{-\frac{\Delta}{2}} & (n-1) \omega & 0 & \displaystyle{\frac{A}{2}} & 0 & 0 & \\
& \displaystyle{-\frac{A}{2}} & 0 & n \omega & \displaystyle{-\frac{\Delta}{2}} & \displaystyle{-\frac{A}{2}} & 0 & \\
& 0 & \displaystyle{\frac{A}{2}} & \displaystyle{-\frac{\Delta}{2}} & n \omega & 0 & \displaystyle{\frac{A}{2}} & \\
& 0 & 0 & \displaystyle{-\frac{A}{2}} & 0 & (n+1) \omega & \displaystyle{-\frac{\Delta}{2}} & \\
& 0 & 0 & 0 & \displaystyle{\frac{A}{2}} & \displaystyle{-\frac{\Delta}{2}} & (n+1) \omega & \\
& & & & & & & \ddots
\end{array}
\right).
\label{Eq:FloquetHamiltonian}
\end{equation}
One can obtain a good approximation for the quasienergies and quasienergy states by numerical diagonalization of a truncated version of $H_F$ as long as the truncated matrix remains sufficiently large. In our calculations, we truncate the matrix $H_F$ with $n$ ranging from $-50$ to $50$.

\subsection*{Choice of quasienergies and quasienergy states}
Although the eigenvalue problem in Eq.~\eq{eigenequation} has an infinite number of solutions, there are only two inequivalent solutions and all the other solutions are copies obtained by shifting an integer number of quanta $n\omega$ between the quasienergy and the periodic part of the Floquet states. We take the two eigenvalues of which difference corresponds to the Rabi frequency $\Omega_\t{R} = |\Delta-\omega|$ in the $A\rightarrow 0$ limit as the two inequivalent solutions. The consideration of evolution in the Floquet picture requires the decomposition of qubit states into quasienergy states at $A=0$ and $t=0$. While any two orthogonal states are a proper choice of Floquet states at $A=0$, we chose $\ket{u_{0,1}}=(\ket{0}\pm\ket{1})/\sqrt{2}$. This choice corresponds to assuming a finite value of $A$ and then taking the limit $A\rightarrow 0$.

\subsection*{Analytical formula for quasienergies}
In this section, we derive an approximate formula for quasienergies of a qubit biased at the symmetry point and subject to harmonic driving. Starting from Eq.~\ref{Eq:FloquetHamiltonian} in the last section, we now perform a basis transformation that physically corresponds to going to a rotating frame (with a time-dependent rotation frequency). The basis states after the transformation are related to those before the transformation by the formula~\cite{S_Shirley:1965sI}
\begin{eqnarray}
\ket{\tilde{u}_{j,n,\uparrow}} & = & \{\dots , J_{-1}\left(\frac{A}{\omega}\right) , 0 , J_0\left(\frac{A}{\omega}\right) , 0 , J_1\left(\frac{A}{\omega}\right) , 0 , \dots \} , \nonumber \\
\ket{\tilde{u}_{j,n,\downarrow}} & = & \{\dots , 0 , J_{-1}\left(-\frac{A}{\omega}\right) , 0 , J_0\left(-\frac{A}{\omega}\right) , 0 , J_1\left(-\frac{A}{\omega}\right) , \dots \}.
\end{eqnarray}
The Hamiltonian for the new basis now reads
\begin{equation}
\tilde H_F = \left(
\begin{array}{cccccccc}
\ddots & & & & & & & \\
& (n-1) \omega & \displaystyle{-\frac{\Delta}{2}J_0\left(\frac{2A}{\omega}\right)} & 0 & \displaystyle{-\frac{\Delta}{2}J_1\left(\frac{2A}{\omega}\right)} & 0 & \displaystyle{-\frac{\Delta}{2}J_2\left(\frac{2A}{\omega}\right)} & \\
& \displaystyle{-\frac{\Delta}{2}J_0\left(\frac{2A}{\omega}\right)} & (n-1) \omega & \displaystyle{\frac{\Delta}{2}J_1\left(\frac{2A}{\omega}\right)} & 0 & \displaystyle{-\frac{\Delta}{2}J_2\left(\frac{2A}{\omega}\right)} & 0 & \\
& 0 & \displaystyle{\frac{\Delta}{2}J_1\left(\frac{2A}{\omega}\right)} & n \omega & \displaystyle{-\frac{\Delta}{2}J_0\left(\frac{2A}{\omega}\right)} & 0 & \displaystyle{-\frac{\Delta}{2}J_1\left(\frac{2A}{\omega}\right)} & \\
& \displaystyle{-\frac{\Delta}{2}J_1\left(\frac{2A}{\omega}\right)} & 0 & \displaystyle{-\frac{\Delta}{2}J_0\left(\frac{2A}{\omega}\right)} & n \omega & \displaystyle{\frac{\Delta}{2}J_1\left(\frac{2A}{\omega}\right)} & 0 & \\
& 0 & \displaystyle{-\frac{\Delta}{2}J_2\left(\frac{2A}{\omega}\right)} & 0 & \displaystyle{\frac{\Delta}{2}J_1\left(\frac{2A}{\omega}\right)} & (n+1) \omega & \displaystyle{-\frac{\Delta}{2}J_0\left(\frac{2A}{\omega}\right)} & \\
& \displaystyle{-\frac{\Delta}{2}J_2\left(\frac{2A}{\omega}\right)} & 0 & \displaystyle{-\frac{\Delta}{2}J_1\left(\frac{2A}{\omega}\right)} & 0 & \displaystyle{-\frac{\Delta}{2}J_0\left(\frac{2A}{\omega}\right)} & (n+1) \omega & \\
& & & & & & & \ddots
\end{array}
\right).
\label{Eq:FloquetHamiltonianBessel}
\end{equation}
It is helpful at this point to use the picture of perturbation theory and think of all the off-diagonal matrix elements as a perturbation. Since there are a large number of matrix elements in the perturbation component and many of these will have negligible effects, we will try to identify the elements that need to be kept for an accurate description of the system. The unperturbed Hamiltonian is given by
\begin{equation}
\tilde H_{F}^{(0)} = \left(
\begin{array}{cccccccc}
\ddots & & & & & & & \\
& (n-1) \omega & 0 & 0 & 0 & 0 & 0 & \\
& 0 & (n-1) \omega & 0 & 0 & 0 & 0 & \\
& 0 & 0 & n \omega & 0 & 0 & 0 & \\
& 0 & 0 & 0 & n \omega & 0 & 0 & \\
& 0 & 0 & 0 & 0 & (n+1) \omega & 0 & \\
& 0 & 0 & 0 & 0 & 0 & (n+1) \omega & \\
& & & & & & & \ddots
\end{array}
\right).
\end{equation}
Let us now calculate the quasienergy and Floquet state starting from the third entry in the matrix above, i.e.~the one that corresponds to the first of two appearances of the unperturbed energy $n\omega$. In the unperturbed Hamiltonian, this state is degenerate with the one that corresponds to the fourth entry. We therefore have to add the part of the perturbation that lifts this degeneracy:
\begin{equation}
\tilde H_F' = \left(
\begin{array}{cccccccc}
\ddots & & & & & & & \\
& (n-1) \omega & 0 & 0 & 0 & 0 & 0 & \\
& 0 & (n-1) \omega & 0 & 0 & 0 & 0 & \\
& 0 & 0 & n \omega & \displaystyle{-\frac{\Delta}{2}J_0\left(\frac{2A}{\omega}\right)} & 0 & 0 & \\
& 0 & 0 & \displaystyle{-\frac{\Delta}{2}J_0\left(\frac{2A}{\omega}\right)} & n \omega & 0 & 0 & \\
& 0 & 0 & 0 & 0 & (n+1) \omega & 0 & \\
& 0 & 0 & 0 & 0 & 0 & (n+1) \omega & \\
& & & & & & & \ddots
\end{array}
\right).
\end{equation}
For large values of $A$ this minimal addition to the Hamiltonian is indeed sufficient to obtain a good approximation for the quasienergies and Floquet states, because when $A\gg\omega$ the Bessel functions are all much smaller than one, and we obtain two quasienergies with a separation given by the well-known expression $\Delta J_0\left(2A/\omega\right)$. A problem arises, however, for small values of $A/\omega$. In the limit $A/\omega\rightarrow 0$, one quasienergy is shifted up (from the unperturbed value) by $\Delta/2$ and the other quasienergy is shifted down by the same amount. When that happens (and assuming that $\omega$ and $\Delta$ are either exactly resonant or near resonance with each other), each one of the two quasienergies becomes degenerate or nearly degenerate with a similarly shifted level coming from the neighbouring pair of energy levels. As a result, if we now focus on the third entry (as above), we need to include the matrix elements in the Hamiltonian that describe mixing with the energy-level pair just above it in the matrix:
\begin{equation}
\tilde H_F'' = \left(
\begin{array}{cccccccc}
\ddots & & & & & & & \\
& (n-1) \omega & \displaystyle{-\frac{\Delta}{2}J_0\left(\frac{2A}{\omega}\right)} & 0 & \displaystyle{-\frac{\Delta}{2}J_1\left(\frac{2A}{\omega}\right)} & 0 & 0 & \\
& \displaystyle{-\frac{\Delta}{2}J_0\left(\frac{2A}{\omega}\right)} & (n-1) \omega & \displaystyle{\frac{\Delta}{2}J_1\left(\frac{2A}{\omega}\right)} & 0 & 0 & 0 & \\
& 0 & \displaystyle{\frac{\Delta}{2}J_1\left(\frac{2A}{\omega}\right)} & n \omega & \displaystyle{-\frac{\Delta}{2}J_0\left(\frac{2A}{\omega}\right)} & 0 & 0 & \\
& \displaystyle{-\frac{\Delta}{2}J_1\left(\frac{2A}{\omega}\right)} & 0 & \displaystyle{-\frac{\Delta}{2}J_0\left(\frac{2A}{\omega}\right)} & n \omega & 0 & 0 & \\
& 0 & 0 & 0 & 0 & (n+1) \omega & 0 & \\
& 0 & 0 & 0 & 0 & 0 & (n+1) \omega & \\
& & & & & & & \ddots
\end{array}
\right).
\end{equation}
In other words, after truncation and (with no loss of generality) choosing $n=0$, we need to diagonalize the matrix
\begin{equation}
\tilde H_{F,\rm truncated} = \left(
\begin{array}{cccc}
- \omega & \displaystyle{-\frac{\Delta}{2}J_0\left(\frac{2A}{\omega}\right)} & 0 & \displaystyle{-\frac{\Delta}{2}J_1\left(\frac{2A}{\omega}\right)} \\
\displaystyle{-\frac{\Delta}{2}J_0\left(\frac{2A}{\omega}\right)} & - \omega & \displaystyle{\frac{\Delta}{2}J_1\left(\frac{2A}{\omega}\right)} & 0 \\
0 & \displaystyle{\frac{\Delta}{2}J_1\left(\frac{2A}{\omega}\right)} & 0 & \displaystyle{-\frac{\Delta}{2}J_0\left(\frac{2A}{\omega}\right)} \\
\displaystyle{-\frac{\Delta}{2}J_1\left(\frac{2A}{\omega}\right)} & 0 & \displaystyle{-\frac{\Delta}{2}J_0\left(\frac{2A}{\omega}\right)} & 0
\end{array}
\right).
\label{Eq:4x4TruncatedHamiltonian}
\end{equation}
We now perform a basis transformation $\tilde{H} = S^{\dagger} \tilde H_{F,\rm truncated} S$, with
\begin{equation}
S = \frac{1}{\sqrt{2}} \left(
\begin{array}{cccc}
1 & 1 & 0 & 0 \\
1 & -1 & 0 & 0 \\
0 & 0 & 1 & 1 \\
0 & 0 & 1 & -1
\end{array}
\right),
\end{equation}
and we obtain
\begin{equation}
\tilde H = \frac{1}{2}
\left(
\begin{array}{cccc}
-2\omega-\displaystyle{\Delta J_0\left(\frac{2A}{\omega}\right)} & 0 & 0 & \displaystyle{\Delta J_1\left(\frac{2A}{\omega}\right)} \\
0 & -2\omega+\displaystyle{\Delta J_0\left(\frac{2A}{\omega}\right)} & \displaystyle{-\Delta J_1\left(\frac{2A}{\omega}\right)} & 0 \\
0 & \displaystyle{-\Delta J_1\left(\frac{2A}{\omega}\right)} & -\displaystyle{\Delta J_0\left(\frac{2A}{\omega}\right)} & 0 \\
\displaystyle{\Delta J_1\left(\frac{2A}{\omega}\right)} & 0 & 0 & \displaystyle{\Delta J_0\left(\frac{2A}{\omega}\right)}
\end{array}
\right).
\end{equation}
This matrix can be split into two $2 \times 2$ decoupled blocks, the relevant one of which is
\begin{equation}
\frac{1}{2}
\left(
\begin{array}{cccc}
-2\omega + \displaystyle{\Delta J_0\left(\frac{2A}{\omega}\right)} & \displaystyle{-\Delta J_1\left(\frac{2A}{\omega}\right)} \\
\displaystyle{-\Delta J_1\left(\frac{2A}{\omega}\right)} & \displaystyle{-\Delta J_0\left(\frac{2A}{\omega}\right)}
\end{array}
\right).
\label{Eq:2x2TruncatedHamiltonian}
\end{equation}
The eigenvalues of this matrix (i.e.~the quasienergies) are given by:
\begin{eqnarray}
\epsilon_0 & = & - \frac{\omega}{2} - \frac{1}{2}\sqrt{ \left[ \omega - \Delta J_0\left(\frac{2A}{\omega}\right) \right]^2 + \Delta^2 J_1^2\left(\frac{2A}{\omega}\right) } \nonumber \\
\epsilon_1 & = & - \frac{\omega}{2} + \frac{1}{2}\sqrt{ \left[ \omega - \Delta J_0\left(\frac{2A}{\omega}\right) \right]^2 + \Delta^2 J_1^2\left(\frac{2A}{\omega}\right) }. \nonumber \\ \label{eq:ApproxQuasienergy}
\end{eqnarray}
The Rabi frequency is given by the difference between the two quasienergies:
\begin{equation}
\Omega_R = \sqrt{ \left[ \omega - \Delta J_0\left(\frac{2A}{\omega}\right) \right]^2 + \Delta^2 J_1^2\left(\frac{2A}{\omega}\right) }.
\label{eq:RabiFrequencyAnalyticalFormula}
\end{equation}
We note here that the expression for the quasienergy difference depends on our choice of quasienergies; choices other than the one used above would give expressions for the Rabi frequency that are different from the one given in Eq.~\eq{RabiFrequencyAnalyticalFormula} by an integer multiple of $\omega$. The choice used here has the advantage that it gives the most natural expression for the Rabi frequency at small values of $A$, especially given the fact that there is only one dominant frequency component in the qubit population dynamics in the weak-driving regime. At higher values of $A$, it becomes more a matter of convention which frequency in the spectrum one defines as the Rabi frequency. In the weak-driving limit $(A\ll\omega)$, this expression reduces to
\be
\Omega_R = \sqrt{ \left( \omega - \Delta \right)^2 +
\Delta^2\left(\frac{A}{\omega}\right)^2 },
\ee
which upon assuming $\Delta\approx\omega$ reduces to the well-known expression
\be
\Omega_R = \sqrt{ \left(\omega - \Delta \right)^2 + A^2 }.
\ee
In the strong-driving limit $(A\gg\omega)$, Eq.~\eq{RabiFrequencyAnalyticalFormula} reduces to
\be
\Omega_R = \omega - \Delta J_0\left(\frac{2A}{\omega}\right),
\ee
which, up to the physically unimportant differences of a shift by $\omega$ and a possible minus sign, is the well-known expression for the Rabi frequency in the strong-driving limit \cite{S_Shirley:1965sI,S_Shevchenko:2010hf}.

\begin{figure}[!]
\includegraphics[width=95mm]{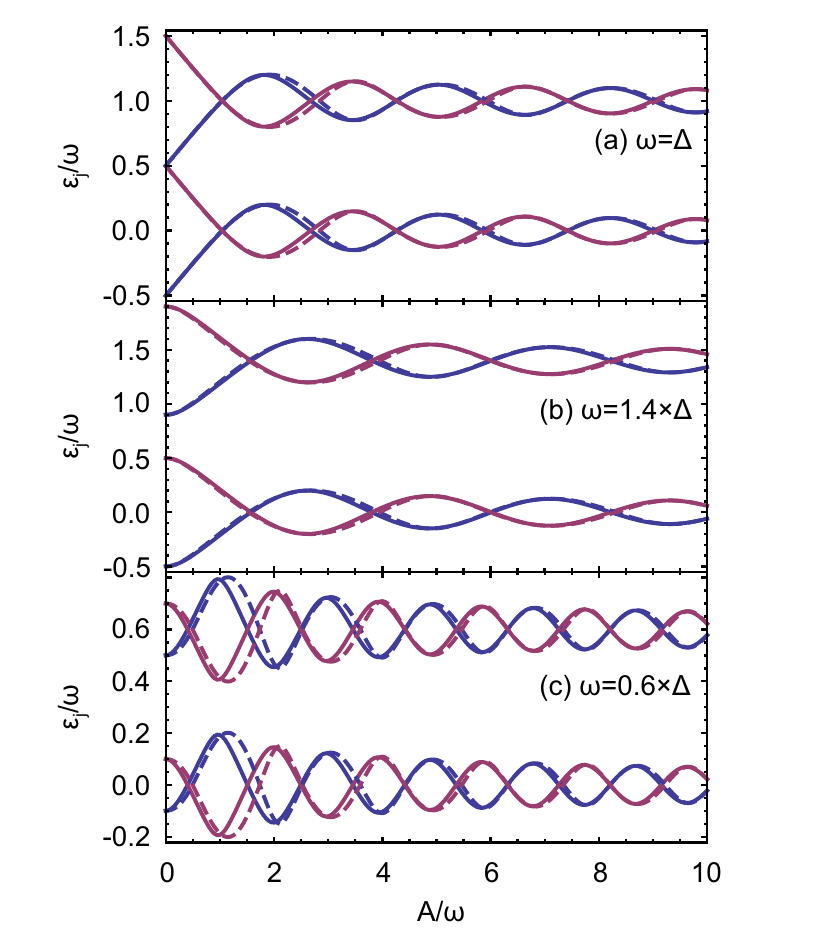}
\caption{\label{sp_fig1} Quasienergies versus the driving amplitude $A$ for on-resonance driving condition, $\omega = \Delta$ (a), and off-resonance driving conditions, $\omega = 1.4\times\Delta$ (b) and $\omega = 0.6\times\Delta$ (c). Solid lines are obtained by the numerical simulations, while dashed lines are obtained from the analytical formula. We show four neighboring quasienergies among the infinite number of solutions. The red and blue lines are quasienergies which correspond to the two inequivalent quasienergies, $\epsilon_0$ and $\epsilon_1$ respectively.}
\end{figure}
In Fig.~\ref{sp_fig1}, we compare quasienergies $\epsilon_{j}(A)$ at different driving frequencies obtained by numerically diagonalizing the Floquet Hamiltonian Eq.~(\ref{Eq:FloquetHamiltonian}) (solid lines) and the analytical formula Eq.~\eq{ApproxQuasienergy} (dotted lines). The numerical simulations use a Floquet Hamiltonian of matrix size of $101\times 101$. The agreement between the numerical simulations and the analytical formula is very good in general. However, small disagreement is visible in the regime where $A\sim \omega$. This is because we ignore many off-diagonal matrix elements in Eq.~(\ref{Eq:FloquetHamiltonianBessel}), which contains Bessel function $J_n(2A/\omega)$ with $n \ge 2$. We note that these Bessel functions vanish at $A\ll\omega$ and $A\gg \omega$ and they have maximum absolute values around $A\sim \omega$.

\end{document}